\documentclass[tighten, twocolumn, trackchanges]{aastex63}
\usepackage{xcolor, fontawesome, microtype, amsmath}
\definecolor{twitterblue}{RGB}{64,153,255}
\definecolor{linkcolor}{rgb}{0.1216,0.4667,0.7059}
\newcommand{\twitter}[1]{\href{https://twitter.com/#1}{\textcolor{twitterblue}{\faTwitter}\,\tt \textcolor{twitterblue}{@#1}}}

\shorttitle{Interstellar communication network. II. Deep space nodes with gravitational lensing}
\shortauthors{Hippke}
\begin{document}
\title{Interstellar communication network.\\II. Deep space nodes with gravitational lensing}

\author[0000-0002-0794-6339]{Michael Hippke}
\affiliation{Sonneberg Observatory, Sternwartestr. 32, 96515 Sonneberg, Germany \twitter{hippke}}
\affiliation{Visiting Scholar, Breakthrough Listen Group, Berkeley SETI Research Center, Astronomy Department, UC Berkeley}
\email{michael@hippke.org}

\begin{abstract}
Data rates in an interstellar communication network suffer from the inverse square law due to the vast distances between the stars. To achieve high (Gbits/s) data rates, some combination of large apertures and high power is required. Alternatively, signals can be focused by the gravitational lenses of stars to yield gains of order $10^{9}$, compared to the direct path. Gravitational lens physics imposes a set of constraints on the sizes and locations of receivers and apertures. These characteristics include the minimum and maximum receiver size, the maximum transmitter size, and the heliocentric receiver distance. Optimal sizes of receivers and transmitters are of order meters. Such small devices allow for the capture of the main lobe in the beam while avoiding the temporal smearing which affects larger apertures. These and other properties can be used to describe the most likely parameters of a lensed communication network, and to determine exact position of communication nodes in the heliocentric reference frame.
\end{abstract}


\section{Introduction}

\begin{figure}
\includegraphics[width=\linewidth]{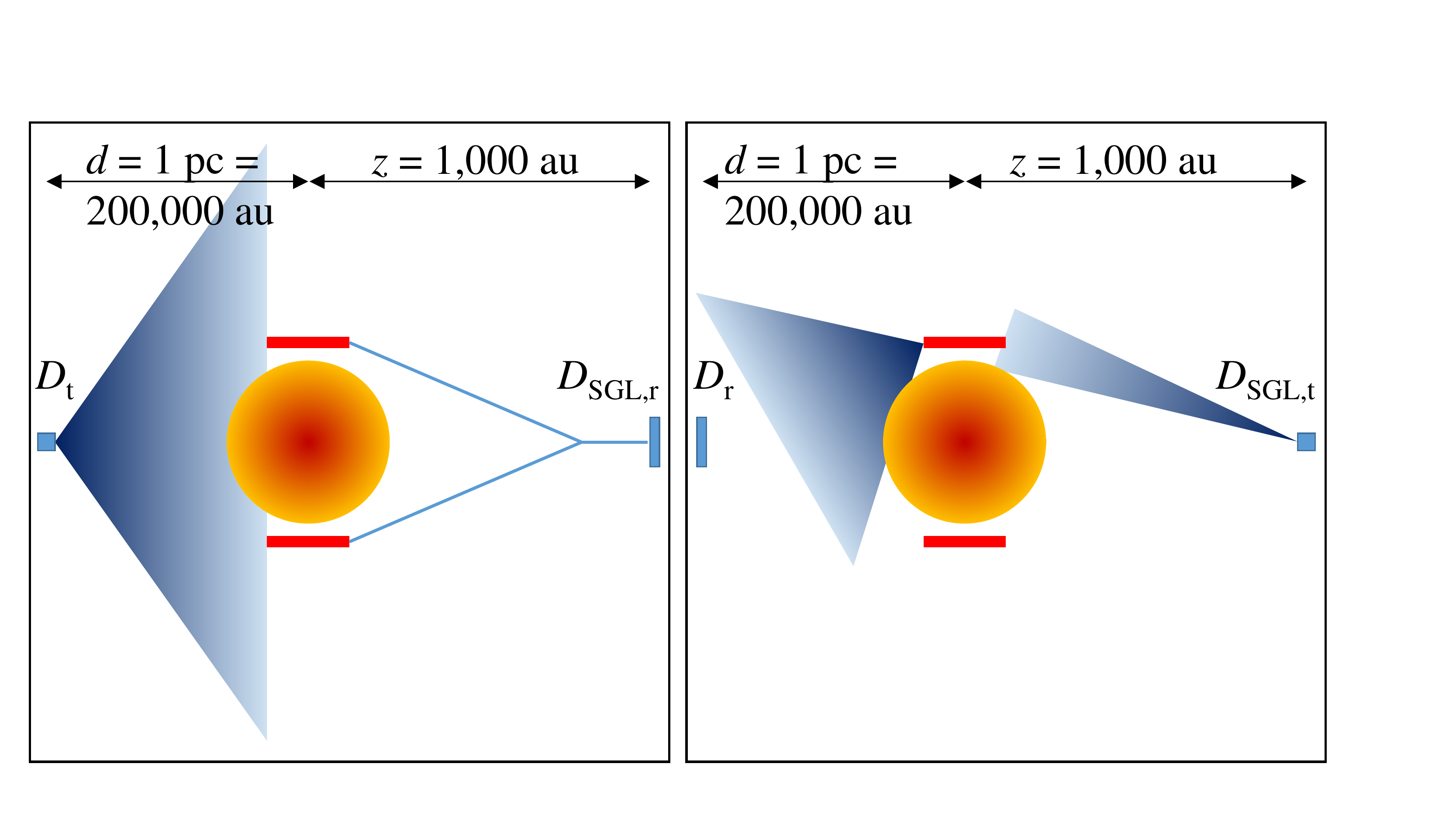}
\caption{Cartoon diagram of SGL configurations. Left: A transmitter at parsec distance beams towards the sun. The receiver is located in the SGL. Right: The transmitter is in the SGL and beams towards the Einstein ring. The receiver is at parsec distance. Nothing is to scale.}
\label{fig:rays}
\end{figure}

The bending of light around the sun was predicted by \citet{1911AnP...340..898E,Einstein1915} and experimentally confirmed during solar eclipses \citep{1919Obs....42..119E,1920RSPTA.220..291D}. Later, \citet{1936Sci....84..506E} also noted the focusing of starlight by the gravitational field of another star. The use of this lensing for communication has been suggested \citep{1979Sci...205.1133E,2011AcAau..68...76M}. While the lens gains are large, it was not clear whether coronal noise would prohibit its use \citep{1981RaSc...16.1473H,Landis2017}. Recent work indicates that a coronograph or occulter enables a stable receiver channel \citep{2018AcAau.142...64H}, as it can block the majority of the noise. A detailed view into the physics of the solar gravitational lens (SGL) can be found in \citet{2017PhRvD..96b4008T}.

A communication scheme of an interstellar network of nodes was introduced in the first paper of this series \citep{2019arXiv191202616H}. It assumes the capability to travel between the stars, and place nodes and probes in multiple locations. Then, it is also feasible to travel to the gravitational lens plane, whose distance is of order $10^{-3}\,$pc from each star. This paper explores the properties of a receiver and a transmitter on the focal line, including their heliocentric locations and sizes. We develop the framework using the more common assumption of a receiver in the lens plane (Figure~\ref{fig:rays}, left panel). Afterwards, we calculate the difference for the case where a transmitter is placed in the same location, while the receiver is at parsec distance in free space (Figure~\ref{fig:rays}, right panel). Scaling relations show that an increase in aperture size for receivers in the SGL produces only a linear increase in data rate, in contrast to a quadratic profit when increasing the transmitter size. These and similar size preferences will be noted, but their consequences for the network topology will be given in a subsequent paper of this series.

\begin{figure*}
\includegraphics[width=\linewidth]{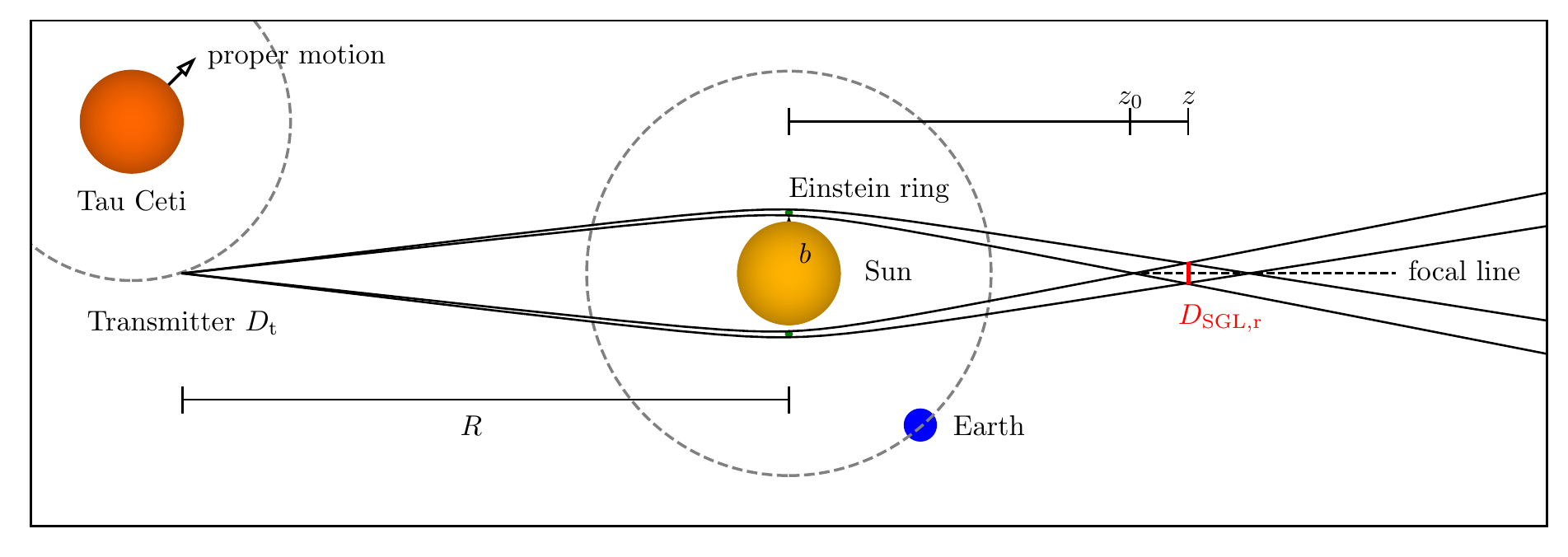}
\caption{Cartoon diagram of the solar gravitational lens configuration. The receiver with aperture $D_{\rm SGL,r}$ on the focal line (right) observes the flux at distance $z$ from the sun which comes through the Einstein ring from the transmitter at distance $d$.}
\label{fig:ray}
\end{figure*}

\section{General lens characteristics}
\label{sec:geometry}
This section describes lens characteristics applicable to both cases of receiver and transmitter in the lens plane. A detailed derivation of these properties can be found in \citet{2017PhRvD..96b4008T}.

\subsection{Schwarzschild radius}
The sun's gravitational field deflects light rays with a bending angle inversely proportional to the impact parameter $b>R_{\odot}$ (Figure~\ref{fig:ray}). This makes it different to a classical lens with its single focal point. In contrast, a gravitational lens has a focal line (a caustic) beginning at a minimum distance $z_{\rm 0}=R_{\odot}^2/2r_g \sim 547.8$\,au, with $r_g = 2GM_{\odot} /c^2 \sim 2,953$\,m as the Schwarzschild radius of the sun \citep{1964PhRv..133..835L,1979Sci...205.1133E}. When nothing else is specified, we assume $z=1{,}000\,$au and a wavelength of $\lambda=1\,\mu$m throughout the paper.

\subsection{Location and width of the Einstein ring}
From the perspective of an observer located in the SGL, the distance between the solar center and the Einstein ring is the impact parameter $b$, which is a function of heliocentric distance $z>z_{\rm 0}$ \citep{2003MNRAS.341..577T}:

\begin{equation}
z(b)=z_{\rm 0}\frac{b^2}{R^2_\odot}
\end{equation}

which can be solved solve for $b$:
\begin{equation}
b(z)=\sqrt{\frac{z}{z_{\rm 0}}}
\label{b(z)}
\end{equation}

in units of $R_{\odot}$. For example, we get $z(b=1.05) \sim 600$\,au and $b(z=1000{\rm~au}) \sim 1.35\,R_{\odot}$. The distance between the solar limb and the Einstein ring is $1-b$.

The area covered by the thin Einstein ring is the effective telescope aperture. This area can be calculated as $A_{\rm ER}=2 \pi ((b + w) ^2 - b^2)$ with $b$ as the inner edge of the ring, $w=D_{\rm SGL,r}$ as the width, and $D_{\rm SGL,r}$ as the aperture of the receiver. A telescope with aperture $D_{\rm SGL,r}$, placed at the heliocentric distance $z$ on the optical axis, receives light from a family of rays with different impact parameters with respect to the Sun, ranging from $b$ to $b+D_{\rm SGL,r}$ \citep[for details, see explanation before Equation 145 in][]{ 2017PhRvD..96b4008T}. A meter-sized telescope sees a meter-sized (width) Einstein ring with a diameter slightly larger than the sun. As the width of the ring is very small compared to the radius ($w/b\approx10^{-9}$) we can approximate $A_{\rm ER}=\pi bw$. The width is unresolved by a telescope in the SGL except for extremely large apertures (section~\ref{sec:max_transmitter_size}).

\subsection{Point spread function}
In the image plane, the flux follows an Airy pattern. The point spread function (PSF) width can be described as \citep[][their Eq.~142]{2017PhRvD..96b4008T}

\begin{equation}
\label{eq:rho}
\rho = 6.3
\left( \frac{\lambda}{1\,\mu{\rm m}} \right)
\left( \frac{z}{1{,}000\,{\rm au}} \right)^{1/2}
{\rm cm}.
\end{equation}

The PSF follows a Bessel function of zeroth order $J_0$, making its decay much slower compared to a classical PSF, which is proportional to $J^2_1(2\sqrt{x})/x^2$. It is therefore possible to scan over the structure of the Airy pattern in the image plane to find the focal line.

\begin{figure}
\includegraphics[width=\linewidth]{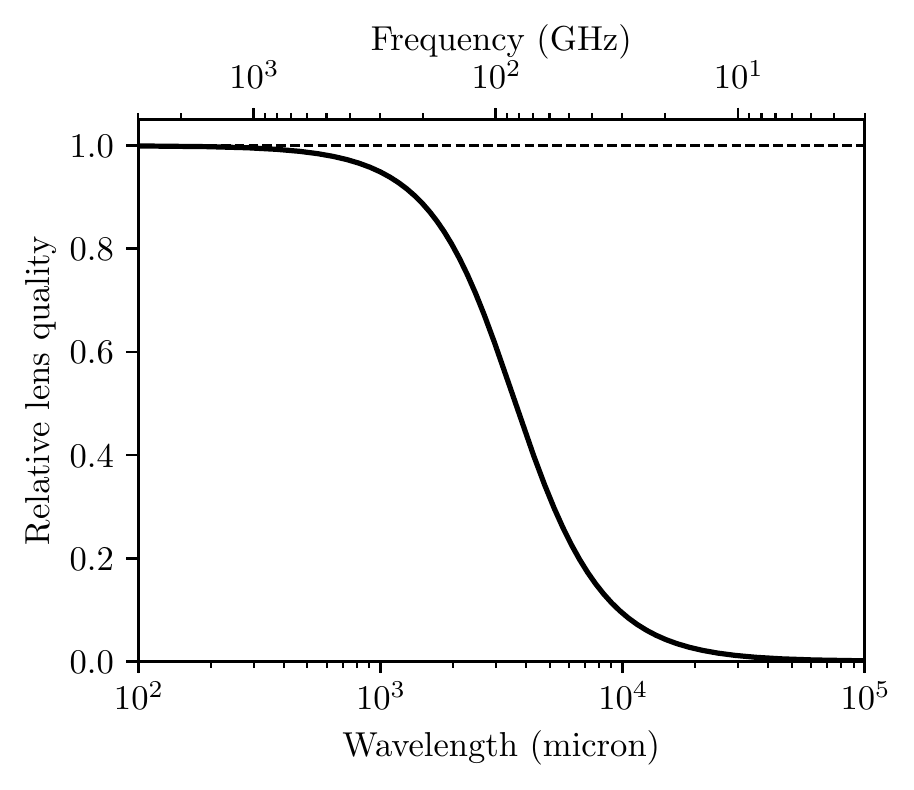}
\caption{Quality of the lens as a function of wavelength. Lens quality is near-perfect for $\lambda\gtrsim100\,\mu$m ($f\lesssim 1{,}000\,$GHz), but deteriorates quickly for lower frequencies. Estimates are from the electron density model following Eq. 285 in \citet{2019PhRvD..99b4044T}.}
\label{fig:lens_quality}
\end{figure}

\subsection{Upper limit on the wavelength}
\label{sec:wavemax}
No lens is perfect, and the same is true for the solar gravitational lens. The turbulent free-electron gas in the sun's atmosphere blocks photons $\lesssim 191\,$MHz \citep{2019EPJP..134...63T}. Higher frequencies are bent towards the sun by plasma refraction, reducing the flux on the focal line. Earlier estimates found a cut-off frequency $f<f_\textrm{crit}\sim123$\,GHz, for which no focus occurs. Later estimates are at 300 GHz \citep{2011JBIS...64...24G}. The most recent model indicates a gradual decrease in lens quality between 10 and $1{,}000\,$GHz \citep{2019PhRvD..99b4044T}. For higher frequencies, i.e. wavelengths shorter than $100\,\mu$m, the effects on the lens are minor (Figure~\ref{fig:lens_quality}).

It has been pointed out by \citet{2018JBIS...71..369L} that the variation in the refractive index produces a randomness in the phase of the focused light, which degrades the diameter of the focal spot considerably. The resulting spot size based on a model by \citet{2005ExA....20..307K} is of order 10\,m. This would make the ``pencil'' beam a rather large brush, and require $D_{\rm SGL}$ of a similar size. Other models by \citet{2019PhRvD..99b4044T,2019EPJP..134...63T,2019JOpt...21d5601T} estimate the bluring effect as smaller than a meter. This is an interesting question to be answered by future solar (or deep space) missions.

\subsection{Lower limit on the wavelength}
\label{sec:wavemin}
Space-based communication does not suffer from atmospheric absorption, but only from interstellar extinction due to gas and dust. Over parsec distances, any wavelength other than the Lyman continuum ($\approx50 \dots 91.2\,$nm) is transparent to $>0.95$ \citep[Figure~2 in][]{2019IJAsB..18..267H}. Shorter wavelengths can be focused more tightly for a given aperture size, making them preferable. This assumes the availability of a competitive transmitter at the desired wavelength, and the technology to focus it. Physical surfaces are limited to nm accuracy due to the finite size of atoms \citep{2012OptEn..51a1013W,2016JOpt...18g4011B,2017arXiv171105761H}. X-ray optics could bypass the limit with Fresnel optics \citep{2004SPIE.5168..459S}, but at the expense of long focal lengths \citep[$10^3\,$km,][]{2012SPIE.8443E..41B}. Pros and cons of various wavelengths will be explored in a future paper of this series.

\section{Receiver in the lens plane}
\label{sec:receiver}
There are two possible configurations to consider. In the first, the receiver is located in the lens plane (Figure~\ref{fig:ray}). The second, with the transmitter in the lens plane, will be treated in section~\ref{sec:trans}.

\subsection{Lens gain}
The gain for distant point sources on-axis is a function of wavelength $\lambda$ and heliocentric distance $z$ \citep[][their Eq. 135]{2017PhRvD..96b4008T}

\begin{eqnarray}
\mu=\frac{4\pi^2}{1-e^{-4\pi^2 r_g/\lambda}}\frac{r_g}{\lambda}\, J^2_0\Big(2\pi\frac{\rho}{\lambda}\sqrt{\frac{2r_g}{z}}\Big)
\end{eqnarray}

The sum of the collected transmitter flux can be calculated by spherically integrating over the PSF \citep[][their Eq. 143]{2017PhRvD..96b4008T} for the receiving telescope aperture in the SGL $D_{\rm SGL,r}$

\begin{eqnarray}
\begin{aligned}
\label{eq_mu}
\bar{\mu}(z,D_{\rm SGL,r},\lambda)=\frac{4\pi^2}{1-e^{-4\pi^2 r_g/\lambda}}\frac{r_g}{\lambda} \times \\ \Big\{ J^2_0\Big(\pi\frac{D_{\rm SGL,r}}{\lambda}\sqrt{\frac{2r_g}{z}}\Big)+J^2_1\Big(\pi\frac{D_{\rm SGL,r}}{\lambda}\sqrt{\frac{2r_g}{z}}\Big)\Big\}.
\end{aligned}
\end{eqnarray}

The corresponding aperture of a standard telescope is $D_{\rm classical} = D_{\rm SGL,r} \sqrt{\bar{\mu}} \sim 87\,$km. This shows the large gain of the SGL as a receiver: an availability of infinitely many ``natural megastructures'' in the form of gigantic telescopes in the vicinity of each star in the galaxy.

For small sources at finite (but very large) distances, we can estimate the lens flux by comparing the power transmission through the lens \citep{2019arXiv190903116T}

\begin{equation}
P_{\rm SGL} =  \frac{d^2}{4 R^2} \frac{\sqrt{2 r_g \overline z}}{D_{\rm SGL,r}}
\end{equation}

where $\overline{z} = z^2/R^2 + z/R + z + 1$ are heliocentric distances along the line between the point source at distance $R$ (typically at least a parsec away) and the center of the Sun. On the direct path, we receive

\begin{equation}
P_0 = \frac{d^2}{16 R^2}
\end{equation}

and obtain a ratio (the lens gain) of

\begin{equation}
\frac{P_{\rm SGL}}{P_0} = \bar{\mu} = 4 \frac{\sqrt{2 r_g \overline z}}{D_{\rm SGL,r}}
\end{equation}

\label{sec:zgain}
We can approximate the influence of $z$ and $D_{\rm SGL,r}$ on $\bar{\mu}$ as

\begin{equation}
\label{eq:gain}
\bar{\mu} \sim 4 \times 10^9
\left( \frac{z}{1000\,{\rm au}} \right)^{1/2}
\left( \frac{D_{\rm SGL,r}}{1\,{\rm m}} \right)^{-1}
\end{equation}

which is valid for $D_{\rm SGL,r} \gg \rho$ and $\lambda \lesssim 1\,\mu$m.

The slow increase of gain with $z$ gives some (but not much) motivation to move the probe further out from the sun. The gain increase with $z$ is true for a coherent light source (such as a laser transmitter), but not when using the lens as a telescope to observe distant planets.

\begin{figure}
\includegraphics[width=\linewidth]{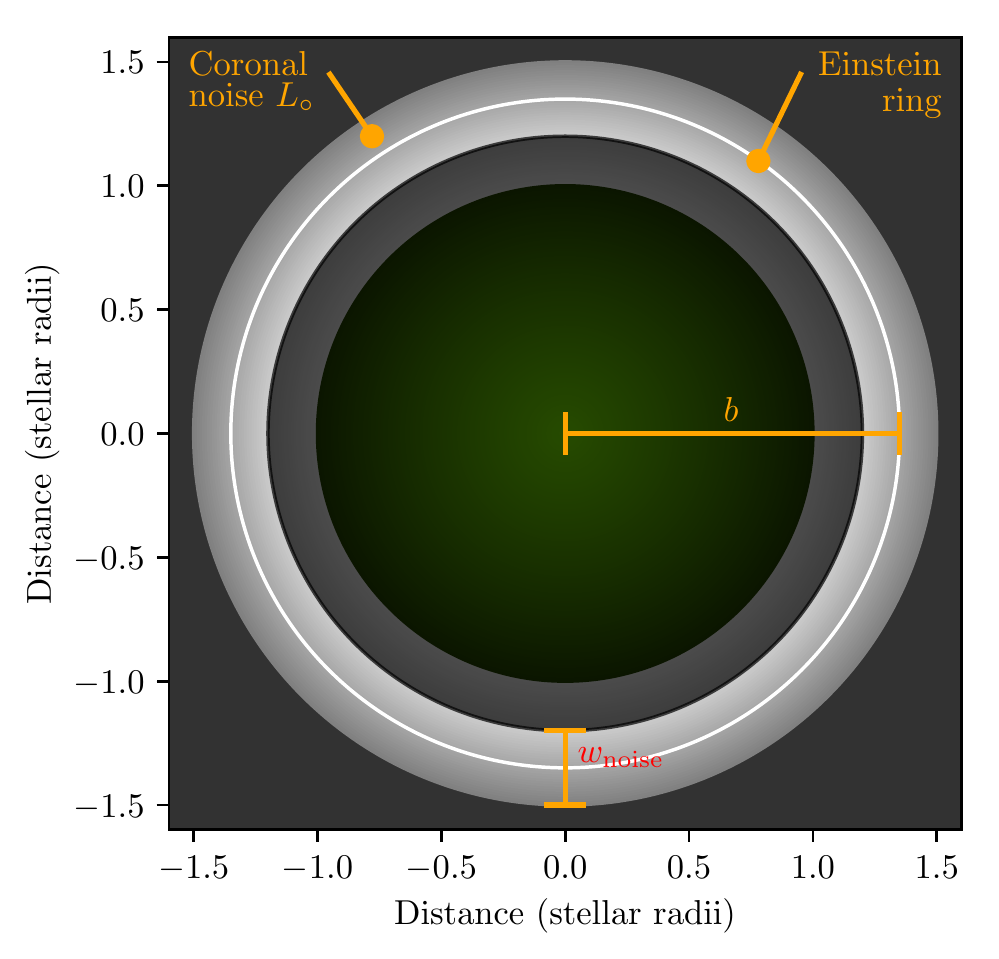}
\caption{Sketch of what the receiver sees through its ring-shaped coronograph from $z=1{,}000\,$au. Flux from the solar disc and the innermost and outermost parts of the corona are strongly suppressed. A ring-shaped area is observed with a width $w_{\rm noise}$. The Einstein ring (white line) is at a distance $b/R_{\odot}=1.35$ from the solar center, in the middle of this area.}
\label{fig:sun}
\end{figure}

\subsection{Minimum receiver size}
\label{sec:min_size}
There are two physical constraints on the minimum aperture size $D_{\rm SGL,r}$ of the telescope onboard the probe in the focal plane. The first comes from the size of the point-spread function, which is 6.3\,cm in the standard scenario ($z=1{,}000\,$au, $\lambda=1\,\mu$m, $D_{\rm SGL,r}=0.1\,$m). Detectors smaller than the PSF would lose a large fraction of the lens gain, making this a first sensible minimum size \citep[see Figure~3 in][]{2018AcAau.142...64H}.

The second, more important minimum comes from the requirement to resolve the gap between the solar limb and the Einstein ring with at least one resolution element \citep[see Fig. 7 in][]{2018AcAau.142...64H}. If the resolution is lower, the ring blends with the sun's disk and is overwhelmed by noise (Figure~\ref{fig:sun}, left). As seen from $z=1000$\,au, the sun subtends 1.91\,arcsec with the Einstein ring located at $b/R_{\odot}=1.35$, surrounded by the corona. The gap between ring and limb is thus $\theta=0.34$ arcsec. It takes a telescope of about 0.75\,m aperture to resolve this gap at $\lambda=1$\,$\mu$m.

More formally, from the perspective of an observer in the SGL, the apparent radius of the sun is $R_{\odot}/z$, giving the distance between the Einstein ring and the solar limb as $(b R_{\odot} - R_{\odot}) / z$. From Eq.~\ref{b(z)} we can substitute $b(z)=\sqrt{z/z_0}$ and require the usual minimum aperture to resolve objects ($1.22\,\lambda/D$), so that

\begin{equation}
D_{\rm SGL,r} > 1.22\,\lambda z  \left(R_{\odot}  \big( \sqrt{z/z_0} - 1 \big)     \right)^{-1}.
\end{equation}

\begin{figure}
\includegraphics[width=\linewidth]{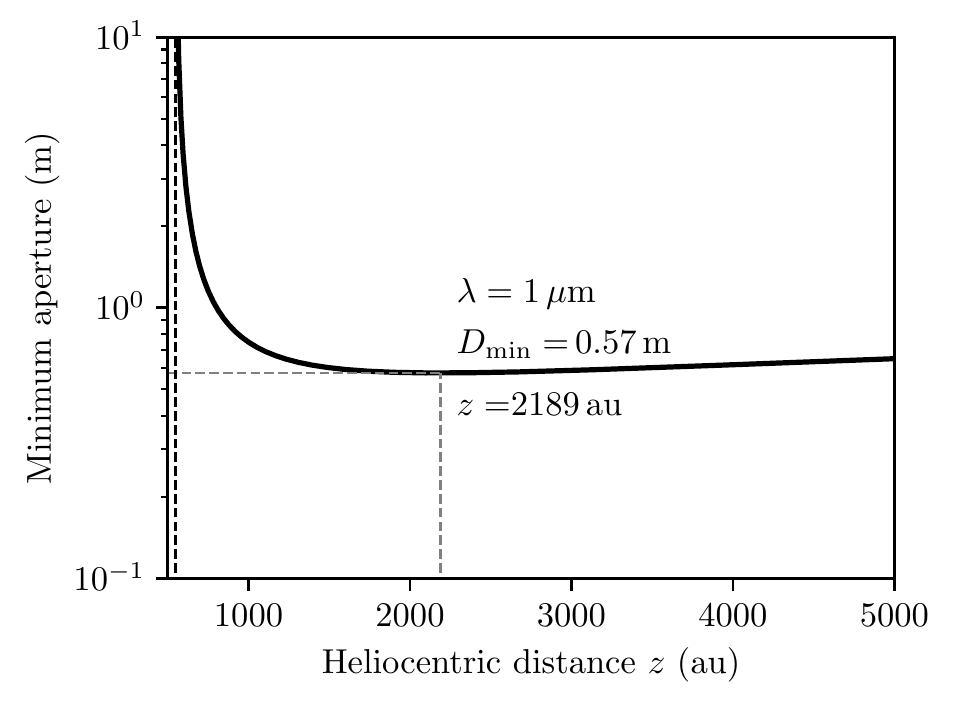}
\caption{Minimum aperture for $\lambda=1\,\mu$m to resolve the gap between the limb and the Einstein ring, a function of heliocentric distance. The minimum is found at $z=2{,}189\,$au independently of wavelength.}
\label{fig:aperture_resolution}
\end{figure}

The function is shown in Figure~\ref{fig:aperture_resolution} with one resolution element to obtain an encircled energy of 84\,\%, as the first null of the airy pattern of the PSF touches the solar limb. An aperture twice as large achieves an energy encirclement of 91\,\%, reducing resolution-induced losses to $<10\,$\%. Depending on the unknown technology of the coronograph used to suppress the sun's light, it is reasonable to assume a minimum aperture with a few times the formal minimum value. Thus, for optical wavelengths, we expect minimum telescope sizes of order meters.

This holds for heliocentric distances $z \gtrsim 1{,}000\,$au. For smaller distances, the required minimum aperture increases strongly, and reaches 2.7\,m at 600\,au. For $1{,}000<z<8{,}000\,$au, we can approximate the minimum aperture (to within 20\,\%) as

\begin{equation}
D_{\rm SGL,r} = 0.6 \left( \frac{\lambda}{1\,\mu{\rm m}} \right) {\rm m}.
\end{equation}

Interestingly, the minimum aperture reaches a global minimum at $z=2{,}189\,$au independent of wavelength. Further out, the inverse-square law dominates over the slowly increasing limb-ring distance.

These results are in agreement with similar calculations required to resolve annuli around the Einstein ring for the purpose of exoplanet imaging \citep[Eq.~96 in][]{2019arXiv190801948T}; although this work treats only the ``flat'' part of the aperture relation for large $z$.

The second minimum (to resolve the gap) is larger than the first (to capture the PSF to its first null) for all useful combinations of $z$ and $\lambda$. A special case would be the usage of an occulter in formation flight \citep[see section~4.4 in][]{2018AcAau.142...64H}. Only the occulter would need to be sufficiently large to occult the gap, while the receiver itself could then only be large enough to catch the PSF. This scenario, however, involves two spacecraft in formation flight, making the scheme much more complex (section~\ref{sec:alignement}).

With the minimum size established, is it preferable to make the receiver larger than this? Probably not, as shown in the next section.

\subsection{Maximum receiver size}
\label{sec:linear}
At first glance, the additional value (in received flux) from larger receiver apertures $D_{\rm SGL,r}$ seems very tempting. For instance, at $\lambda=1\,\mu$m we have a gain of $\bar{\mu}=4\times10^9$, and the corresponding classical aperture for $D_{\rm SGL,r}=1$\,m is $D_{\rm classical} = D_{\rm SGL,r} \sqrt{\bar{\mu}} \sim 87\,$km (section~\ref{sec:geometry}). For $D_{\rm SGL,r}=2\,$m, this grows to  $D_{\rm classical}=123\,$km. While this looks impressive, the scaling relation of flux to the equivalent classical aperture diameter is only linear.

For comparison, in classical telescopes diffraction limits the beam to an angle of $\theta \sim \lambda / D$. An increase in aperture diameter $D$ leads to a linear shrinkage of the beam width, and thus to a quadratic increase in beam intensity per surface area. A linear increase in aperture thus increases photon flux rates quadratically, all else being equal. A light bucket of aperture $D_{\rm r}$ collects photons of an area $\pi (D_{\rm r}/2)^2$. Again, a linear increase with aperture yields a quadratic increase in photon flux.

So why does the SGL receiver scale only linearly with $D_{\rm SGL,r}$? The answer is in the effective aperture as given by the area of the Einstein ring, $A_{\rm ER}=\pi bw$. Doubling the receiver aperture doubles the width of the Einstein ring. For $D_{\rm SGL,r} \ll R_{\odot}$ this also ``only'' doubles the area of the ring. Equivalently, the aperture is not filled but ring-shaped; it has a very large central obstruction preventing the quadratic increase of classical telescopes.

\subsection{Maximum transmitter size}
\label{sec:maxtrans}
Naively, if we start with $D_{\rm SGL,r}=D_{\rm t}=1\,$m and we could add aperture size at will, we would gain most by giving all increase to the transmitter. However, we can not increase $D_{\rm t}$ beyond all bounds due to the lensing geometry. When the transmitter beam becomes so tight that it is smaller than the star at arrival, almost all flux would be lost into the star; no lensing would occur.

Alternatively, the transmitter could center its beam on the Einstein ring instead of the stellar midpoint, assuming that the transmitter knows the receiver's position and thus the location of $b$ (at the time of arrival of the photons). Then, the lens would form an Einstein arc \citep[see section~6 in][]{2018AcAau.142...64H}, which is preferable for beam widths tighter than $3R_{\odot}$. This approach is sound from a lensing perspective, but it creates temporal distortions. The lens is now not symmetric any more (with the sun at its center), but asymmetric (centered on the ring). Some rays travel a path up a solar radius longer than others. Part of the flux arrives later in the receiver, delayed by the extra light travel time $\Delta t \sim 5\,$s required to traverse $\leq 2R_{\odot}$. An unaffected channel could leverage $M = f \sim 10^{14}$ temporal encoding modes per photon (in a $t_{\rm dur}=1\,$s transmission). The smear reduces the number of temporal encoding modes to one per 5\,s. Yet, spectral encoding (as well as polarization) should remain unaffected. In any case, communication with high data rates becomes either impossible or extremely challenging with such strong temporal smearing.

With a minimum beam width of $3R_{\odot}$ at arrival, the maximum transmitter apertures (assuming diffraction limited optics) can be calculated. Over a parsec distance, the beam width limit is equivalent to $D_{\rm t}\sim150\,$m at $\lambda=1\,\mu$m, or $D_{\rm t}=1\,$m at $\lambda\sim6.7\,$nm, or generally

\begin{equation}
D_{\rm t,\,max} = 150
\left( \frac{\lambda}{1\,\mu{\rm m}} \right)
\left( \frac{d}{1\,{\rm pc}} \right)
{\rm m}.
\end{equation}

Of course, telescope apertures have other cost factors, such as the cost of construction and maintenance. The human cost-aperture relationship is as least quadratic \citep{2004SPIE.5489..563V}, but it is unknown for future or non-human technology.

A second limit to the maximum transmitter aperture comes from the fact that the resolution of the SGL is very high. In other words, the plate scale in the focal plane is very large. From geometrical optics, we wish to keep $D_{\rm SGL,r} > D_{\rm t} (z/d)$ where $d$ is the distance to the transmitter (parsecs away). Intuitively, if the transmitter is too large, its virtual image in the image plane is larger than the receiver aperture, and part of the lensed flux would be lost. In other words, the width $w$ of the Einstein ring is too large to be captured by the receiver aperture. In physical units, we require

\begin{equation}
D_{\rm t,\,max} = 200
\left( \frac{D_{\rm SGL,r}}{1\,{\rm m}} \right)
\left( \frac{z}{1{,}000\,{\rm au}} \right)
\left( \frac{d}{1\,{\rm pc}} \right)^{-1}
{\rm m}.
\end{equation}

In practice, this issue makes exoplanet imaging difficult, because the receiver will be much smaller than the virtual image, and must scan over this (moving, rotating) image of km size. For communication, the limit is less problematic. Of course, this limit on $D_{\rm t,\,max}$ could be inverted to pose a limit on the minimum $D_{\rm SGL,r}$ instead, for a given $D_{\rm t}$.

\subsection{Receiver size penalty from temporal smearing}
\label{sec:distort}
From Figure~\ref{fig:ray} it is clear that the width of the Einstein ring is small but finite. Rays at $b+w$ have a longer path compared to rays that traverse through $b$. The outer rays arrive later by $\sqrt{2} w$, causing a temporal smearing of $\Delta t = \sqrt{2} D_{\rm SGL,r} \, c^{-1}$ seconds, or

\begin{equation}
\label{eq:smearing}
\Delta t =
5 \times 10^{-9}
\left( \frac{D_{\rm SGL,r}}{1\,{\rm m}} \right)
{\rm s}
\end{equation}

independent of wavelength and heliocentric position. The lens is achromatic for $\lambda \lesssim 100\,\mu$m if the aperture is sufficiently large \citep[a few times $\rho$,][]{2019arXiv190903116T}.

The effect that photons transmitted at the same time arrive at different times in the lens plane has also been described before. In \citet[][Equation 30]{2019PhRvD.100h4018T} this extra path length is derived as $\delta r=(b^2/2r_g)^2/z_0$.

Rays following the longer path are affected by less gravitational time dilation, making them faster. The effect is described by Einstein's field equations,

\begin{equation}
\label{eq:einstein}
t_0 = t_f \sqrt{1-\frac{2GM}{rc^2}}
\end{equation}

where $G$ is the gravitational constant, $M=M_{\odot}$ the solar mass, $c$ the speed of light, and $r = b \geq R_{\odot}$ the distance from the center of the gravitational mass. In our case, for rays just grazing the solar limb, the time dilation is a few times $10^{-6}$. The integral over the total path is of order $10^{-5}\,$s. Rays with an impact parameter $b+w$ are dilated less by $\approx10^{-15}$, for a total temporal difference of $<10^{-14}\,$s. Compared to the difference in path length, the dilation effect is smaller by five orders of magnitude, and can thus be neglected.

\subsection{Direct path transmission by the SGL receiver}
With the receiver in the lens plane, direct path communication can be initiated by moving one of the two spacecrafts away from the direct line of the sun. On the heliocentric axis, the transmitter would need to move away more than the apparent size of the sun,

\begin{eqnarray}
\left( \frac{a}{1\,{\rm au}} \right) \left( \frac{d}{1\,{\rm pc}} \right)^{-1}
> \left( \frac{z}{1900\,{\rm au}} \right)^{-1}
\end{eqnarray}

If the distant transmitter is in a 1\,au orbit around its star, a (steady) receiver in the lens plane at $z=1{,}900\,$au would (just) see the transmitter during aphelion.

\section{Transmitter in the lens plane}
\label{sec:trans}
It has been suggested to use two stars as a ``radio bridge'' to double the gain and communicate in both directions \citep[Figure~2 in][]{2014AcAau.104..458M}. This does not work. First, the gravitational lens is unusable in the radio band (section~\ref{sec:wavemax}). More importantly, the PSFs are so small that using only one lens at a time is sufficient.

Instead, we now consider the case where the probe in the SGL wishes to transmit back \textit{through the solar lens} to a location parsecs away. That is, we keep the positions of the object previously called ``receiver'' (for now, the ``transmitter''), but reverse the signal flow. We begin by calculating useful size ranges for transmitters by considering the limiting cases of being able to inject almost all, and essentially none, of their flux into the Einstein ring.

\begin{figure}
\includegraphics[width=.49\linewidth]{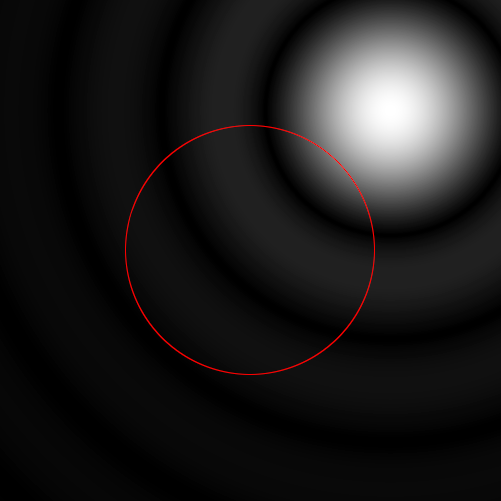}
\includegraphics[width=.49\linewidth]{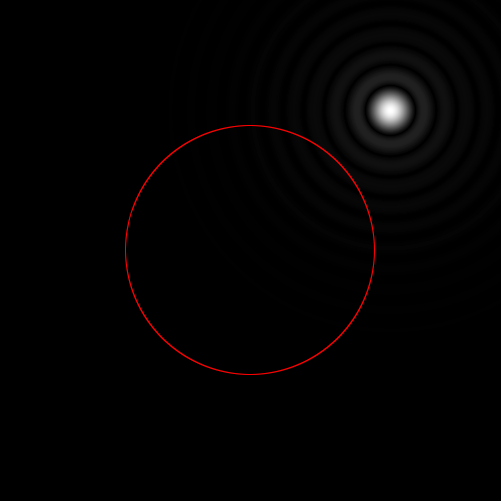}

\includegraphics[width=.49\linewidth]{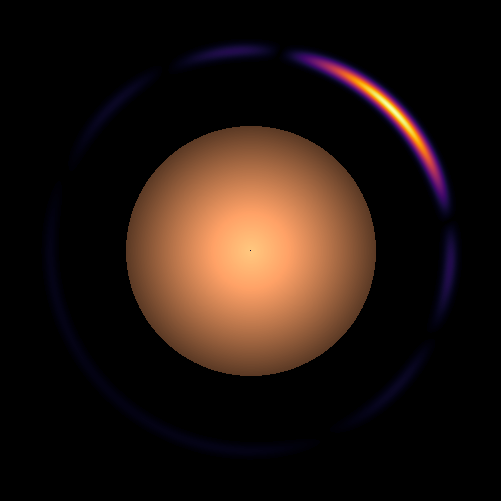}
\includegraphics[width=.49\linewidth]{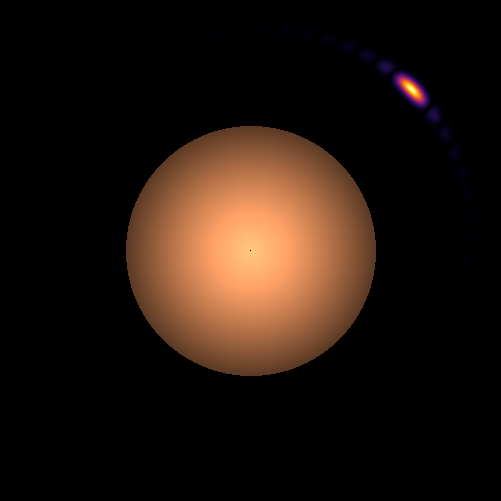}
\caption{Top panels: Beam as seen by an observer at the position of the Einstein ring. The position of the sun is indicated by the red circle. Bottom panels: Einstein ring produced by these beams as seen by a distant observer. Left: Beam by a small telescope ($D_{\rm SGL,t}=0.1\,$m) from $z=1{,}400\,$au producing a large Einstein arc at $b\sim1.6$. Right: Beam by a larger telescope ($D_{\rm SGL,t}=1\,$m) from $z=2{,}200\,$au producing a small Einstein arc at $b\sim2$.}
\label{fig:beams_inverse}
\end{figure}

\subsection{Maximum transmitter size}
\label{sec:max_transmitter_size}
The SGL transmitter injects all flux (to its first null) into the Einstein ring when its beam size is tighter than the width of the ring, i.e. when the diameter of its aperture equals its beam width at the distance to the ring, or

\begin{equation}
\label{eq:max_transmitter_size1}
D_{\rm SGL,t} = \frac{1.22 z \lambda}{D_{\rm SGL,t}}\\
\Longleftrightarrow
D_{\rm SGL,t} = \sqrt{1.22 z \lambda}
\end{equation}

which is $\sim 13.5\,{\rm km}$ for $\lambda=1\,\mu$m and $z=1{,}000\,$au. Generally, the transmitter size saturates at

\begin{eqnarray}
\begin{aligned}
\label{eq:max_transmitter_size2}
D_{\rm SGL,t} \leq 13.5
\left(\frac{z}{1000\,{\rm au}} \right)^{-1}
\left(\frac{\lambda}{1\,\mu{\rm m}} \right)^{-1}
\,{\rm km}.
\end{aligned}
\end{eqnarray}

\subsection{Minimum transmitter size}
\label{sec:min_transmitter_size}
There is no hard limit for the minimum size of the transmitter; even an isotropic emitter will deposit some flux into the Einstein ring if it is located in the SGL. For large enough apertures (or short enough wavelengths)

\begin{eqnarray}
\begin{aligned}
D_{\rm SGL,t} \geq 13 \left(\frac{z}{1000\,{\rm au}} \right)^{-1} \left(\frac{\lambda}{1\,\mu{\rm m}} \right)^{-1} \,{\rm cm}
\end{aligned}
\end{eqnarray}

the beam is smaller than the diameter of the sun, so that the main lobe is not lost into the star, but a larger fraction of it may enter the ring.

\subsection{Heliocentric distance}
Reducing the transmitter position from $z$ towards $z_0$ increases the received flux linearly and not quadratically, as one would expect. On the one hand, the beam shape by the transmitter in the SGL towards the Einstein ring follows the classical inverse-square law, which would indicate a quadratic relation. Shorter heliocentric distances $z$ (at constant aperture and wavelength) also cause the beam to illuminate a larger arc length of the Einstein ring. The area inside the ring, however, increases only linearly because of its fixed width $w=D_{\rm SGL,t}$. This holds for the regime where the beam width at arrival is smaller than the ring diameter, i.e. $\lesssim13.5\,$km.

\subsection{Signal gain}
As a useful approximation, the gain for a transmitter in the lens plane is roughly given by the ratio of the area of the active section of the Einstein ring to the area of the transmitting aperture. This is the same principle as outlined in section~\ref{sec:receiver}.

\subsection{Noise}
\label{sec:noise2}
The transmitter is not affected by coronal noise. Instead, the receiver faces the issue that the Einstein ring is at a small sky-projected angle from the star, because it is now at parsec distance from the ring (and not ``only'' $\sim1{,}000\,$au). Starlight and Einstein ring are resolved by apertures

\begin{eqnarray}
\begin{aligned}
D_r > 140\,
\left(\sqrt{\frac{z}{550{\,\rm au}}} -1 \right)^{-1}
\left(\frac{\lambda}{1\,\mu{\rm m}} \right)^{-1}
\left(\frac{d}{1\,{\rm pc}}\right)
\,{\rm m}
\end{aligned}
\end{eqnarray}

which is about 150\,m for $z=1{,}000\,$au over a parsec distance. If both sources can be spatially resolved, noise levels will be low. Small receivers, however, will observe star and ring blended. In this case, the broadband isotropic (noise) flux from a main sequence star can be approximated as

\begin{equation}
N_{\odot} \sim 3 \times10^{10}
\left(\frac{L}{L_{\odot}}\right)
\left(\frac{d}{1\,{\rm pc}}\right)^{-2}
\,{\rm s}^{-1} \,{\rm m}^{-2}.
\end{equation}




\subsection{Temporal smearing}
\label{sec:smear2}
The segment of the Einstein ring which the transmitter illuminates is quite large, typically a fraction of the solar radius. It thus seems as this large area would cause a large temporal smearing effect larger than the one described with Equation~\ref{eq:smearing}. If the transmitter is on axis, however, this is not the case. The time difference between early and late rays is still only equivalent to its aperture size; or equivalently, to the width, and not the length, of the Einstein ring. Intuitively, we could employ the doughnut-shaped beam which would illuminate all of the ring evenly, thus causing only a small smearing. The conical shape produced by the circular beam yields the same result of small temporal smear.

\section{Data rate}
The data rate of the SGL channel is a function of signal photons, noise photons, and bandwidth. Again, we first treat the standard scheme of an SGL receiver.

\subsection{Photon flux for the SGL receiver}
For reference, on the direct path between two telescopes we expect the number of photons as \citep{2017arXiv171105761H}

\begin{eqnarray}
\begin{aligned}
\label{eq:direct}
F_{\rm direct} = 10^{-3}
&\left(\frac{d}{1\,{\rm pc}} \right)^{-2}
\left(\frac{\lambda}{1\,\mu{\rm m}} \right)^{-1}\\
&\left(\frac{D_{\rm r}}{1\,{\rm m}} \right)^{2}
\left(\frac{D_{\rm t}}{1\,{\rm m}} \right)^{2}
\left(\frac{P}{1\,{\rm W}} \right)
\,{\rm s^{-1}}.
\end{aligned}
\end{eqnarray}

where $D_{\rm t}$ is the aperture of a standard transmitter with power $P$, collected by a standard receiving telescope of size $D_{\rm r}$. Both receiver and transmitter aperture have a quadratic influence on the direct path photon flux.

In contrast, through the SGL we get

\begin{eqnarray}
\begin{aligned}
\label{eq:signal_sgl}
F_{\rm SGL} = 3.76 \times 10^6
&\left( \frac{z}{1000\,{\rm au}} \right)^{1/2}
 \left(\frac{d}{1\,{\rm pc}} \right)^{-2}
 \left(\frac{\lambda}{1\,\mu{\rm m}} \right)^{-1} \\
&\left(\frac{D_{\rm SGL,r}}{1\,{\rm m}} \right)
 \left(\frac{D_{\rm t}}{1\,{\rm m}} \right)^{2}
 \left(\frac{P}{1\,{\rm W}} \right)
 \,{\rm s^{-1}}.
\end{aligned}
\end{eqnarray}

The advantage of the SGL link over the direct path can be determined for $D_r=D_{\rm SGL,r}$ as

\begin{eqnarray}
\begin{aligned}
\label{eq:signal_photon_sgl}
\frac{F_{\rm SGL}}{F_{\rm direct}} = 3.76 \times 10^9
\left( \frac{z}{1000\,{\rm au}} \right)^{1/2}
\left(\frac{D_r}{1\,{\rm m}} \right)^{-1}
\end{aligned}
\end{eqnarray}

which is equivalent to Equation~\ref{eq:gain}, as expected.

The data rate (or data per energy) scales as $F_{\rm SGL} \propto D_{\rm SGL,r}$ but $F_{\rm direct} \propto D_{\rm r}^2$. Again, this shows the preference to grow the transmitter aperture rather than the receiver aperture.

These estimate neglect losses from extinction, which are negligible in almost any case over short (parsec) interstellar distances (section~\ref{sec:wavemin}).

\subsection{Noise for the SGL receiver}
\label{sec:noise}
The sun is surrounded by a corona of light reflected off dust particles and spectral emission produced by ions in the plasma. Coronal noise dominates over all other sources of noise by orders of magnitude. Secondary noise sources, such as thermal noise in the receiver or background flux from other astrophysical sources, contribute typically not more than a few photons per second per square meter. The receiver in the SGL can employ a ring-shaped coronograph, which blocks all light except the region covering the Einstein ring. Diffraction will determine the width $w_{\rm noise}$ of this slit and thus its area and luminosity $L_{\odot}$ (Figure~\ref{fig:sun}). The broadband coronal noise can be estimated as \citep{2019IJAsB..18..267H}

\begin{equation}
\label{eq:noise}
N_{\rm SGL} = 10^9
\left( \frac{z}{1000\,{\rm au}} \right)^{-3.5}
\left( \frac{D_{\rm SGL,r}}{1\,{\rm m}} \right)^{-1}
 \,{\rm s^{-1}}.
\end{equation}

The strong decline of noise with $z$ comes from the fact that the flux from the diffraction-limited ring-shaped area $L_{\circ}$ decreases with $b^{-6}$ (or $z^{-3.5}$) due to the fast coronal brightness decrease with radial distance \citep[][Eq. 13]{2018AcAau.142...64H}. These estimates are in agreement with \citet{2018AcAau.152..408W}. Effectively, noise decreases by one order of magnitude between $1{,}000$ and $2{,}000\,$au.

Noise is also a function of receiver aperture. With increasing aperture $D_{\rm SGL,r}$ and thus resolution, the coronal noise ring surrounding the Einstein ring becomes thinner. With increasing aperture, resolution grows, and this ring-shaped area shrinks linearly.

Finally, noise is a function of wavelength, following the coronal solar spectrum (which closely resembles the solar spectrum). For a receiver centered at $\lambda=1\,\mu$m with 100\,\% bandwidth, of order unity of the coronal noise flux is collected. Small bandwidths collect much less noise. For example, pulsed lasers have line widths of 350\,MHz $=3\times10^{-4}$\,nm \citep{1999ApOpt..38.6347D}. Matched narrow-band filters would collect only $\lesssim 100$ photons per second at $z=2{,}000\,$au. A standard nanosecond photon counter would collect on average $10^{-7}$ photons per time slot, making the channel essentially noise-free. The cost of this scheme would be a (comparably) narrow bandwidth, restricting the number of possible bits per photon. As explained in the next section, large bandwidth (and high noise) beat small bandwidth (and small noise) -- assuming technology beyond our own.

\begin{figure*}
\includegraphics[width=.5\linewidth]{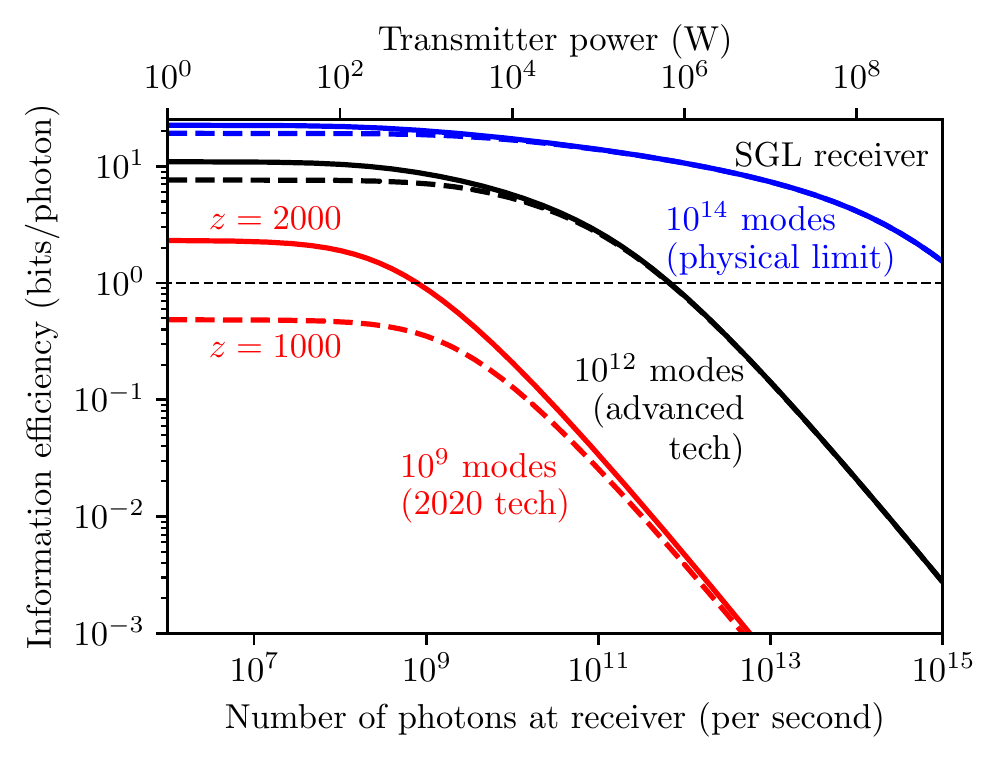}
\includegraphics[width=.5\linewidth]{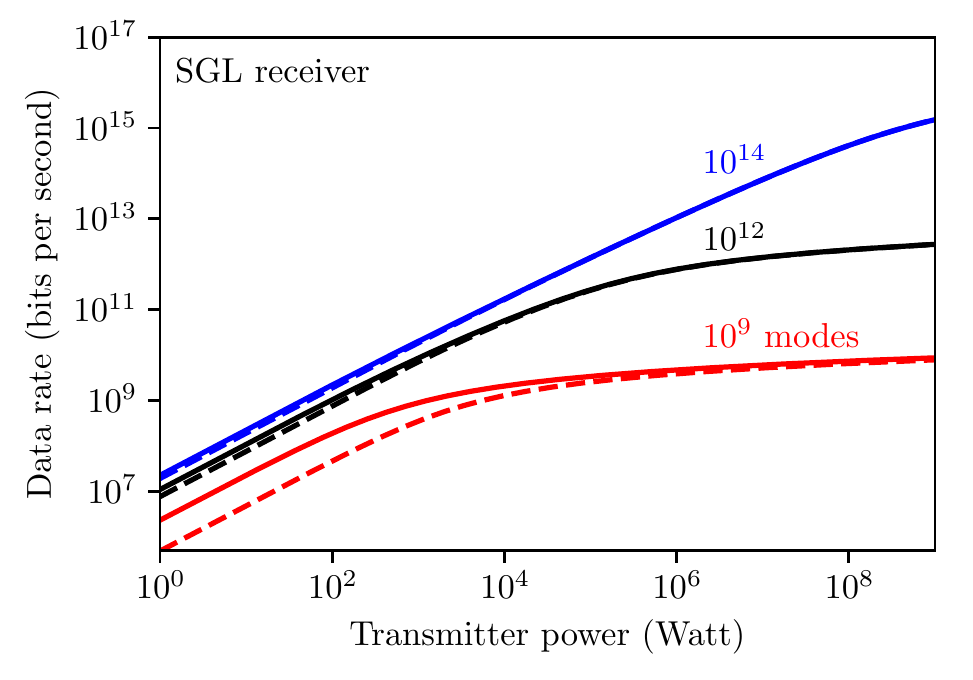}
\caption{SGL receiver scheme. Left: Information efficiency (in bits per photon) as a function of the number of photons received. Solid lines are for receivers at $z=2{,}000\,$au, dashed lines at $z=1{,}000\,$au, suffering higher noise levels. Current technology (red lines) with nanosecond photon counters saturates near $10^9$ photons per second, and can achieve at most a few bits per photon. The limits expand for intermediate (black) and ultimate (blue) technology. Right: Data rates (bit per photon times the number of photons) as a function of transmitter power. In this log-log plot, differences between today's and the ultimate technology are 1--2 orders of magnitude for moderate ($\lesssim \,$kW) power levels.}
\label{fig:pie}
\end{figure*}

\subsection{Encoding modes with temporal smearing}
Information bits can be encoded on the modes of photons, which are energy (i.e. wavelength/frequency), time of arrival, and polarization. Orbital angular momentum is not an independent characteristic as it it broadens the spectrum.

During a one second transmission, the maximum number of modes $M$ at 100\,\% bandwidth and $\lambda=1\,\mu$m is $\Delta t \Delta f \sim 0.5$, or $f \sim M \sim 10^{14}$. Temporal smearing (Equation~\ref{eq:smearing}) reduces the number of time encoding modes to

\begin{equation}
M_{\rm smear} =
2 \times 10^{8}
\left( \frac{D_{\rm SGL,r}}{1\,{\rm m}} \right)
\left( \frac{\lambda}{1\,\mu{\rm m}} \right)^{-1}
{\rm s}^{-1}.
\end{equation}

so that $M/M_{\rm smear} \approx 10^6$ for all $\lambda$. In other words, the remaining $10^6$ modes can only be lifted using spectral encoding. This combination of temporal and spectral modes due to temporal smearing makes ultra-fast (femtosecond) photon counters unnecessary for an optimal receiver. A laser pulse with $\lambda_0=1\,\mu$m and a duration of $\Delta t=5 \times 10^{-9}\,$s has a minimum bandwidth of
$\Delta \lambda = 0.5 \lambda_0^2 \, c^{-1} \Delta t^{-1} \sim 3\times10^{-13}\,$m (0.3\,pm), or about 88 MHz.
The total required bandwidth of $10^6$ spectral modes is 500\,nm, or exactly 50\,\% of $\lambda_0$.

The number of modes increases with smaller $\lambda$. In addition, the aperture $D_{\rm SGL,r}$ can shrink linearly with wavelength, which reduces smearing linearly. As an example, for $\lambda=1\,$nm we would have the number of modes as $M \sim f \sim10^{17}$ and the equivalent aperture $D_{\rm SGL,r}=0.1\,$cm. An issue with such short wavelengths is receiver alignment (section~\ref{sec:alignement}).

\subsection{Number of bits per photon for the SGL receiver}
\label{sec:snr1}
Theoretically, free-space optical communication has unbounded photon information efficiency \citep{2011arXiv1104.2643D}. In practice, limitations arise from finite bandwidth, non-zero noise, and losses. The upper limit to the amount of information which can be transmitted with a quantum state has been established by the \citet{holevo1973bounds} bound for the noiseless case and by \citet{2004PhRvA..70c2315G,2014NaPho...8..796G} with noise, finding a channel capacity of

\begin{equation}
\label{thermal_holevo}
C_{\rm th}=g(\eta M + (1-\eta) N_{\rm M}) - g((1-\eta)N_{\rm M})
\end{equation}

with $\eta$ as the receiver efficiency and $g(x)=(1+x) \log_2 (1+x)-x \log_2 x$ so that $g(x)$ is a function of $\eta \times M$ \citep{2014PhRvA..89d2309T}. The channel uses $M$ modes and collects on average $N_{\rm M}$ noise photons per mode. This limit is higher than the classical \citet{S1949} limit, and has not yet been achieved in practice \citep{2012arXiv1202.0518W,2012SPIE.8246E..05E}. Classical communication can achieve a factor of a few less than the limits discussed here. We will estimate the quantum limits first and compare it to classical limits in section~\ref{sec:holevo}.

Each mode can be ``filled'' with (infinitely many) signal photons, although using more than one photon per mode results in only a logarithmic increase of photon information efficiency (PIE), while the energy requirement increases linearly. To build intuition, consider a quantum receiver which can read $10^9$ modes per second at 100\,\% bandwidth. We assume that the receiver collects $10^9$ signal photons during the transmission, so that the number of (signal) photons per mode is $M=1$. We assume $10^8$ noise photons ($S/N=10$) per second, so that the number of noise photons per mode $N_M=0.1$. With a receiver efficiency of $\eta=0.5$, we can calculate $C_{\rm th}=2.3$ bits per photon from Eq.~\ref{thermal_holevo} (with quantum measurements). The strongest limit on PIE comes from the issue that the modes are well filled with signal photons; adding more bits per photon requires ``stacking'' them, which only gives a logarithmic increase in information density from a linear increase in power, making this approach very energy inefficient. If the receiver could be upgraded to $10^{10}$ modes, we would have $M=0.1$ and $N_M=0.01$, giving $C_{\rm th}=3.9$ bits per photon. The limit here is bandwidth (more precisely, the upper frequency limit).

With an upper limit of $\lambda=1\,\mu$m, it is therefore not worthwhile to push more than $10^{14}$ photons per second to the receiver. What is more, it is not possible to obtain much higher data rates than $10^{14}$ bits per second on a $1\,\mu$m channel, irrespective of the energy usage.

Similarly, we can calculate data rates taking our signal levels from Eq.~\ref{eq:signal_sgl}, noise levels from Eq.~\ref{eq:noise}, and three examples of receiver technology. Here, we consider a quantum receiver which might be possible to build soon with today's technology, using nanosecond photon detectors ($M=10^9$). As a further improvement we investigate a receiver with $M=10^{12}$, perhaps based on nanosecond photon counters which can distinguish $10^3$ colors. Finally, we determine the ultimate limit of $10^{14}$ modes. We test two different noise levels, those present at $z=1{,}000\,$au and $z=2{,}000\,$au with fluxes of $10^9$ and $10^8$ photons per second per square meter. We use a constant aperture of $D_{\rm SGL,r}=1\,$m.

The photon information efficiency in these scenarios varies between of order one bit per photon in the worst case (today's technology at $z=1{,}000\,$au) and $\sim22$ bit per photon in the most favourable case (Figure~\ref{fig:pie}, top left). Intermediate scenarios may achieve of order 10 bits per photon. We consider these the most realistic in practice, as we have neglected all other issues involved, such as misalignment of the receiver with respect to the optical axis, losses inside the receiver, (small) interstellar extinction, etc. In any case, a value between 1 and 20 bits per photon are certainly plausible in this transmission scheme; values outside this span are not expected. The finite number of modes imposes a strong upper limit on the data rate of the channel. As mentioned in the example, $10^9$ modes will not allow to transmit much more than $10^9$ bits per second.

We can take the gain from the lens (Eq.~\ref{eq:signal_sgl}) to determine the relation between transmitter power (in Watt) and data rate (in bits per second) for the scenario examined (Figure~\ref{fig:pie}, top right). A gigabit link ($10^9$ bits per second) is feasible at order of kW power in all scenarios. Terabit links ($10^{12}$) are impossible with nanosecond photon counters and require at least advanced technologies, plus high (MW) powers. The upper limit is found at $10^{15}$ bits per second with ultimate technology, requiring GW transmitter power. These relations of data rate with energy show that an interstellar communication network using nodes at parsec distances does not require (or benefit from) extreme power levels, as would \citet{1964SvA.....8..217K} scale isotropic beacons. In the SGL receiver scheme, energy requirements in the conservative case of $10^9$ modes are $\sim 0.1\,$kJ per GB, or $\sim500\,$GB per kWh.

\subsection{Number of bits per photon for the SGL transmitter}
\label{sec:snr2}
In the same framework it is not straightforward to estimate the data rate and information efficiency for the inverse case where the transmitter is located in the SGL. Temporal smearing and flux gains remain similar (section~\ref{sec:smear2}), while noise (section~\ref{sec:noise2}) increase. Noise levels are of order $10^{10}$ photons per square meter per second, when assuming the star to be blend with the Einstein arc.

As a result of higher noise levels, photon efficiencies are lower. At least advanced technology using $10^{12}$ modes is required to achieve ${\rm PIE}>1$. For current technology, PIE is two orders of magnitude smaller.

\subsection{Quantum versus classical communication}
\label{sec:holevo}
The photon efficiencies calculated in the previous section assumed quantum receivers which are not available today. What makes these receivers different to today's classical receivers (such as photon counters) and how large is the difference on PIE?

Classical receivers measure the arrival (and perhaps energy and polarization) of each photon separately, one-by-one. After the detection of each photon, it is ``removed'' from the channel. Quantum communication, in contrast, requires a receiver which jointly detects long modulated codewords with nonlinear quantum operations. This is considered impossible with current technology. On the other hand, it is known that modulated coherent (laser) light is sufficient to achieve the Holevo capacity \citep{2011PhRvL.106x0502G,2014PhRvA..89d2309T,2018PhRvA..97a2315C}. Quantum receivers could be build with quantum pulse gating \citep{2011OExpr..1913770E,2015PhRvX...5d1017B,2017NatCo...814288A} and ultrashort sub-picosecond pulses \citep{2017OExpr..2512952R}, non-Gaussian elements such as Kerr interactions, interactions with non-Gaussian states \citep{2014PhRvA..89d2309T}, or other, yet unknown methods.

In the regime of low background noise combined with a low data rate, simple schemes such as on-off keying (OOK) and pulse-position modulation (PPM) achieve 99\,\% of the channel capacity \citep{2018arXiv180107778L,2019SPIE10910E..0AB}. In the regime of high background and high data rate (which is our SGL case), the situation is more complex. In the case of ${\rm PIE} \gtrsim 3$ photons per bit, most known classical modulation schemes become inefficient. The gap in energy efficiency between OOK or PPM and the Holevo limit grows linearly with PIE and approaches a factor of 25 in energy at PIE=10 bits per photon.

Interestingly, if a quantum receiver can be build, we can make a strong prediction for the encoding scheme that will be used for our high PIE communication case. It has been proven that the maximum capacity can be achieved with binary phase-shift keying (BPSK) in combination with polar coding \citep{2011arXiv1109.2591W,2012arXiv1202.0533G}. This recently developed code provides a linear block error correction which has been proven to achieve the channel capacity for symmetric binary input at modest $\mathcal{O}(n \log n)$ encoding and decoding complexity \citep{Arikan:2009:CPM:1669561.1669570}. The combination of a ``simple'' quantum receiver plus software error correction as the encoding used to achieve capacity is attractive, because the high latency over interstellar distances (of at least years) makes forward error correction highly beneficial. Integrating error correction as a software-layer for optimal encoding kills two birds with one stone and is computationally efficient.

Listening into a quantum communication with a classical receiver, however, would collect all the photons, but not their meaning. It could prove the presence of a message, but it would not be possible to decode it, as only a fraction (of order 10\,\%) of the content could be inferred. Interestingly, Earth 2020 technology is at most a few decades away from being able to build a quantum SGL node. At present, flying out to 1{,}000\,au and deploying a quantum receiver together with a kW (or MW) laser is perhaps already within reach of a Manhattan-type project.

\section{Discussion}

\subsection{Lens gain for other objects}

Gravitational lenses of other stars have different characteristics. For example, the Schwarzschild radius of Proxima Cen is $r_{g,\,{\rm Cen}}\sim363.3\,$m, resulting in $z_{0,\,{\rm Cen}} \sim 106\,$au, about $5\times$ closer to the star compared to our sun. The gain for Proxima is about a third of our sun. Generally, the gain scales with the square root of mass,

\begin{equation}
\label{eq:gain_with_r_g}
\bar{\mu} \sim 3.76 \times 10^9
\left( \frac{M}{M_{\odot}} \right)^{1/2}.
\end{equation}

This scaling makes lensing similarly effective within an order of magnitude for all stars $0.1<M<10\,_{\odot}$. For the central black hole in our galaxy with a mass of $6\times10^6\,M_{\odot}$, the gain is about $10^{13}$. Using planets as lenses is problematic due to their large values of $z_0$. Even for Jupiter, $z_0=6{,}100\,$au; for the Earth it is $\sim0.1\,$pc.

The pencil-sharp beam of a lensed transmitter widens from 6\,cm at $1{,}000\,$au to 0.9\,m at parsec distance, and further to 100\,m at 8\,kpc (galactic center), or 9\,km at 800\,kpc (Andromeda galaxy). In this scheme, lensed communications out to Gpc distances have a better energy efficiency than direct path links to even the nearest stars (for the same aperture sizes). They come at the cost of very high pointing accuracy requirements.

A related possibility is to use Earth's atmospheric refraction (not gravity) to bend light \citep{2019PASP..131k4503K}. The maximum on-axis gain is $\sim 75{,}000$ for distances of $600{,}000\,$km with a meter-sized aperture. The equivalent classical telescope aperture is $\sim80\,$km, similar to the SGL receiver. Used as a receiver, noise levels require a detailed analysis. For a transmitter in the lens plane, it is presently unclear what shape and extent the PSF has at large (parsec) distances. Further analysis of such lenses is encouraged, and could be extended to larger atmospheric planets such as Jupiter.

\subsection{Cost and complexity of station keeping}
\label{sec:alignement}
We assume that a transmitter is in orbit around a distant star. The image scale in the focal plane of the SGL is $z/d$. An au-wide orbit at 1.3\,pc seen from $z=1{,}000\,$au gives $z/d \sim 260$, corresponding to a receiver orbit in the SGL of 1/260\,au, or $r \sim 6 \times 10^5\,$km, about $1.5\times$ the distance between the Earth and the moon. Such an ellipse needs to be flown in sync by transmitter and receiver. The size of the ellipse scales linearly with $z$ and $1/d$. A linear increase in semimajor axis results in a linear increase of required $\Delta v$, i.e. propulsion, because the ellipse circumference is of order $2 \pi r$.

The required accuracy of this ellipse increases linearly with $\lambda$ and slowly with $z^{1/2}$ due to the shrinkage of the PSF. To capture most of the flux of the PSF, the positional accuracy $a$ (in the lens plane, i.e. orthogonal to the heliocentric axis) should be less than its width,

\begin{eqnarray}
a \lesssim 0.1
\left( \frac{\lambda}{1\,\mu{\rm m}} \right)^{-1}
\left( \frac{z}{1{,}000\,{\rm au}} \right)^{-1/2}
{\rm m}
\end{eqnarray}

A positional accuracy of $a\sim0.1\,$m during formation flight appears feasible with Earth 2020 technology. At $\lambda=1\,$nm (X-rays), however, this requirement would be $a\sim100\,\mu$m, which might be impossible even for advanced technology. Gravitational displacements of the transmitter could result in a misalignment. The plate scale factor of $z/d \sim 260$ is independent of wavelength. There will be some minimum wavelength for which available technology will pose a limit in alignment.

\subsection{Alignment for SGL receivers}
To initiate a communication, the transmitter should start to transmit at the highest possible power level while the receiver scans space to find the narrow beam. We now estimate the difficulty of this task.

We assume that the transmitter is located at 1\,au from its host star at a distance of a parsec, a fact which we assume the receiver to know. We also assume that the receiver knows of the transmitter's ephemeris around the distant star. It is useful for the receiver to first determine the location of the stellar focus, i.e. the point where the other star is in the focus of the lens plane, which is trivial to find due to the high stellar luminosity. The transmitter focus is then located orthogonal to this point at a distance of $z/d \sim 6 \times 10^5\,$km at pc distance, neglecting inclination. The center of the PSF has a width of 6.3\,cm, or $1/10^{10}$ of the distance between the stellar and the transmitter foci. Determining such an absolute length is trivially possible with laser ranging. For comparison, lunar laser ranging can achieve mm accuracy over $10\times$ longer distances.

With the orthogonal offset, the required ellipsoidal track is determined. The transmitter crosses this virtual track once a year, so that the apparent velocity is $2 \pi (z/d)\,$yr$^{-1}\sim430\,$km\,h$^{-1}$, about the speed of a racing car. Depending on how well the ephemeris is known, the transmitter will scan a small fraction of this track until the signal is detected. The scan process to achieve ${\rm SNR}=10$ at a flux level of $10^9$ photons per second requires to observe for one second at each meter of the track. A full ellipse scan would require 120 years. If the positional ephemeris (the position of the transmitter on its ellipse) is known to 1\,\%, the scan time-to-sync is about a year, which is short compared to interstellar travel times.

\begin{table*}
\center
\caption{Limits for the SGL receiver scheme and the corresponding transmitter
\label{tab:limits1}}
\begin{tabular}{llll}
\tableline
Section & Topic & Limit &  Comment \\
\tableline
\ref{sec:min_size} & Min receiver size & $D_{\rm SGL,\,min} = 1.22\,\lambda z  \left(R_{\odot}  \big( \sqrt{z/z_0} - 1 \big)     \right)^{-1}$ & Resolve gap between solar limb/Einstein ring \\
& & \textcolor{white}{$D_{\rm SGL,\,min}$} $= 0.6 \left( \frac{\lambda}{1\,\mu{\rm m}} \right) {\rm m}$ & for $1{,}000<z<8{,}000\,$au \\
\ref{sec:min_size} &  & $D_{\rm SGL,\,min} = 0.063
\left( \frac{\lambda}{1\,\mu{\rm m}} \right)
\left( \frac{z}{1{,}000\,{\rm au}} \right)^{1/2}
{\rm m}$ & Capture PSF to first null \\
\ref{sec:geometry}; \ref{sec:distort} &  & $D_{\rm SGL,\,min} = 0.3
\left( \frac{\lambda}{1\,\mu{\rm m}} \right)
\left( \frac{z}{1{,}000\,{\rm au}} \right)^{1/2}
{\rm m}$ & Keep lens achromatic\\
\ref{sec:distort} & Max receiver size & $\Delta t =
5 \times 10^{-9}
\left( \frac{1\,{\rm m}}{D_{\rm SGL,r}} \right)
{\rm s}$ & Minimize temporal smearing\\
\ref{sec:linear} &  & $F_{\rm SGL,\,signal} \propto D_{\rm SGL,r}$ & Received flux grows only linearly with the\\
& & &  receiver, but quadratically with the transmitter\\
\ref{sec:maxtrans} & Max transmitter size & $D_{\rm t,\,max} = 150
\left( \frac{\lambda}{1\,\mu{\rm m}} \right)
\left( \frac{d}{1\,{\rm pc}} \right)
{\rm m}$ & Beamwidth $<3R_{\odot}$ to avoid Einstein arcs\\
\ref{sec:maxtrans} & & $D_{\rm t,\,max} = 200
\left( \frac{D_{\rm SGL,r}}{1\,{\rm m}} \right)
\left( \frac{z}{1{,}000\,{\rm au}} \right)
\left( \frac{d}{1\,{\rm pc}} \right)^{-1}
{\rm m}$ & Capture the lensed flux \\
\ref{sec:zgain} & Min heliocentric distance & $\mu \propto z^{1/2};\,\,z_0>547.8\,$au & Slow increase in receiver gain with $z$ \\
\ref{sec:noise} & & $N \propto z^{-3.5}$ & Coronal noise decreases strongly with $z$ \\
\ref{sec:alignement} &  & $a \lesssim 0.1
\left( \frac{\lambda}{1\,\mu{\rm m}} \right)^{-1}
\left( \frac{z}{1{,}000\,{\rm au}} \right)^{-1/2}
{\rm m}$ & Alignment more difficult for small $\lambda$, large $z$ \\
\tableline
\end{tabular}
\end{table*}

\begin{table*}
\center
\caption{Limits for the SGL transmitter scheme and the corresponding receiver
\label{tab:limits2}}
\begin{tabular}{llll}
\tableline
Section & Topic & Limit &  Comment \\
\tableline
\ref{sec:distort} & Max transmitter size & $\Delta t =
5 \times 10^{-9}
\left( \frac{1\,{\rm m}}{D_{\rm SGL,t}} \right)
{\rm s}$ & Minimize temporal smearing\\
\ref{sec:max_transmitter_size} & & $D_{\rm SGL,t} \leq 13.5 \left(\frac{z}{1000\,{\rm au}} \right)^{-1} \left(\frac{\lambda}{1\,\mu{\rm m}} \right)^{-1} \,{\rm km}$ & Inject all flux into Einstein ring segment \\
\ref{sec:min_transmitter_size} & Min transmitter size & $D_{\rm SGL,t} \geq 13 \left(\frac{z}{1000\,{\rm au}} \right)^{-1} \left(\frac{\lambda}{1\,\mu{\rm m}} \right)^{-1} \,{\rm cm}$ & Illuminate ring segment instead of full sun (soft limit) \\
\ref{sec:zgain} & Min heliocentric distance & $F_{\odot} \propto z^{-1};\,\,z_0>547.8\,$au & Shorter distance to the Einstein ring \\
\tableline
\end{tabular}
\end{table*}

\begin{table*}
\center
\caption{Limits which apply to both schemes
\label{tab:limits3}}
\begin{tabular}{llll}
\tableline
Section & Topic & Limit &  Comment \\
\tableline
\ref{sec:wavemax} & Max wavelength & $\lambda_{\rm max} \approx 100\,\mu$m  & Free electron plasma effect \\
\ref{sec:wavemin} & Min wavelength & $\lambda_{\rm min} \approx 1\,$nm  & Atomic surface smoothness limit\\
\tableline
\end{tabular}
\end{table*}

\section{Summary and conclusion}

\subsection{SGL receiver}
In this paper we have learned that a receiving telescope in the SGL can harvest a gain of order $10^9$ if its aperture is $D_{\rm SGL,r} \gtrsim\,1\,$m for $\lambda=1\,\mu$m. This limit comes from the requirement to resolve the gap between the solar limb and the Einstein ring, and is valid for heliocentric distances $1{,}000<z<8{,}000\,$au. Two similar limits of the same order are obtained from the requirement to capture (at least) the flux from the PSF to its first null, and to keep the lens achromatic (Table~\ref{tab:limits1}).

Increasing the receiver side beyond (order of) this limit is unattractive for two reasons. First, the lensed flux undergoes a temporal smearing which scales with the width of the Einstein ring, and thus linearly with $D_{\rm SGL,r}$. It affects the number of encoding modes and is not critical for 1\,m sized receivers, but limits temporal resolution to $\sim50\,$ns. Second, data rates scale only linearly with increases in $D_{\rm SGL,r}$. The root cause is the source of the gain, which is the surface area of the Einstein ring, and which suffers from the large central obstruction caused by the sun. Instead, increasing the transmitter yields an increase in data rate which is quadratic with its aperture.

Increasing the transmitter has similar limitations, although at larger scale. For $\lambda=1\,\mu$m, apertures larger than 150\,m produce beam widths at arrival approaching the solar diameter, making lensing impractical. The alternative, namely focusing on the Einstein ring to produce arcs, is impractical because of strong (order 5\,s) temporal smearing. The second limit comes from the high resolution of the lens in the image plane, or in other words, the large plate scale. To capture at least the PSF to its first null, the transmitter to receiver size ratio should be less than about 200 per parsec of distance.

Regarding the heliocentric distance, it is preferable to position the receiver at $z>z_0$, but not at extremely large distances. The gain increase with distance is mediocre and scales as $\mu \propto z^{1/2}$. Aperture requirements decrease until and increase beyond the optimum position at $z=2{,}189\,$au to resolve the gap between limb and Einstein ring. Beyond this distance, data rates grow only with the square root of the distance, but alignment becomes linearly more difficult. Finally, communications to the inner solar system become quadratically more difficult, due to the inverse square law.

\subsection{SGL transmitter}
A useful transmitter sitting in the SGL will be at least large enough to resolve the gap between the sun and the Einstein ring (13\,cm at micron wavelength), and at most 13.5\,km large as this is sufficient to resolve the ring. Temporal smearing will again limit the maximum size, plausible to meter levels. Some limits apply to both schemes, such as the minimum and maximum wavelengths (Table~\ref{tab:limits3}).

\subsection{The grand scheme}
The lensing scheme presented here is about communications between nodes. In every link between nodes, one probe must be located in the SGL, while the other can be anywhere in free space. One of the nodes must constantly perform position corrections to make the link continuous over time. If this is done on the SGL side, the required $\Delta v$ is low because the plate scale is $z/d$.

SGL nodes are at $1{,}000\,$au from each star, which is $\approx 0.01\,$pc, or 1\,\% of the typical distance between stars in our part of the galaxy. This distance is too far away from many targets of interests, such as planets and moons, which are in the inner stellar systems; usually within a few au from the star. The scheme at hand therefore requires additional, separate \textit{exploration probes} in the inner solar system to gather intelligence and surveillance. The explorers would transmit their surveillance data over a free space link to an SGL transmitter in the same system. Following Equation~\ref{eq:direct}, a direct path transmission with $D_{\rm t}=D_{\rm SGL,t}=1\,$m over $1{,}000\,$au can only provide less than 1\,Mbits/s at kW power and $\lambda=1\,\mu$m, or less than Gbits/s with an X-ray laser. This is less than the lensed link, but still a useful amount of information. The $D_{\rm SGL,t}$ node receives this information and relays it (transfers it) over the lensed link to a nearby stellar system. Alternatively, exploration probes can aim their data at an SGL receiver in a nearby star system. This way, the lensing gain can be leveraged to increase energy efficiency (bits per Joule) by order of $10^5$ (when comparing 1\,pc to $1{,}000\,$au).

\subsection{Future work}
The analysis of the physical properties of transmitters and receivers in the gravitational lens give insights into their technological requirements and most likely characteristics. If we assume that the laws of physics are equally valid for all species in the universe, we could use size and location preferences to steer searches for artifact SETI.

Further work on related limitations and preferences is encouraged. Some aspects might be more important than others in a real application. Some aspects might be surpassed by future insights. Only through refinements and corrections, we will learn how to build a communication network, or how it would be build by other civilizations.

The exact numbers of the tools involved may change by perhaps a factor of a few, depending on the validity of the theoretical models. But it is clear that extremely small \citep{barrow1999impossibility} or extremely large (``alien megastructure'') receivers and transmitters will not work in this gravitational lensing communication scheme.

With these descriptions of communications tool at hand, plus their locations in the heliocentric reference frame, we can proceed to calculate the apparent lens locations on our night sky, as seen from the Earth, with arc-second accuracy. This will be the focus of the next paper in this series.
\\
\\
\textit{Acknowledgments} I thank Slava Turyshev for many helpful discussions.
\bibliography{references}
\end{document}